\documentclass{elsart}

\usepackage{epsfig}

\begin{document}

\begin{frontmatter}

\title{Contact angle hysteresis of cylindrical drops on chemically heterogeneous striped surfaces}


\author{Masao Iwamatsu\thanksref{label2}}
\thanks[label2]{ E-mail: iwamatsu@ph.ns.musashi-tech.ac.jp, Tel: +81-3-3703-3111 ext.2382, 
Fax: +81-3-5707-2222}
\address{Department of Physics, General Education Center,
Musashi Institute of Technology,
Setagaya-ku, Tokyo 158-8557, Japan }

\begin{abstract}
Contact angle hysteresis of a macroscopic droplet on a heterogeneous 
but flat substrate is studied using the interface displacement model. 
First, the apparent contact angle of a droplet on a heterogeneous 
surface under the condition of constant volume is considered.  By assuming 
a cylindrical liquid-vapor surface (meniscus) and minimizing the total 
free energy, we derive an equation for the apparent contact angle, which 
is similar but different from the well-known Cassie's law.  Next, using 
this modified Cassie's law as a guide to predict the behavior of a 
droplet on a heterogeneous striped surface, we examine several scenarios 
of contact angle hysteresis using a periodically striped surface 
model. By changing the volume of the droplet, we predict a sudden 
jump of the droplet edge, and a continuous change of the apparent contact 
angle at the edge of two stripes.  Our results suggest that as drop 
volume is increased (advancing contact lines), the predominant drop 
configuration observed is
the one whose contact angle is large; whereas, decreasing drop volume 
from a large value (receding contact lines) yields drop configuration 
that  predominantly exhibit the smaller contact angle.  
\end{abstract}

\end{frontmatter}

\section{Introduction}
Contact angle hysteresis on a real surface has been investigated experimentally for many years~\cite{Butt,Dettre}.  Usually a higher advancing contact angle and a lower receding contact angle are observed.  This hysteresis has been attributed either to the roughness~\cite{Butt,Shuttleworth,Eick,Huh} or to the chemical heterogeneity~\cite{Johnson,Neumann,Joanny,Marmur,Brandon,Drelich,Swain}. In particular, chemical heterogeneity is most important for smooth surface.  It also plays a crucial role for the morphology and motion of a small droplet~\cite{Chen} on a smooth surface which has attracted much attention recently for its potential use in microfluidic devices~\cite{Lipowsky,Oron,Quere}. 

On a chemically heterogeneous surface, Cassie's law~\cite{Cassie2} for the apparent contact angle~\cite{Young} of a macroscopic droplet is adopted for the design issue of the microfluidic devices in chemical and biomedical engineering.  For example, this law is successfully used to explain the superhydrophobicity and self-cleaning mechanism of various natural and artificial surfaces~\cite{Quere}. This Cassie's law states that the apparent contact angle of the droplet is given by the aerial average of the local contact angles about the droplet base.  Therefore, the contact angle hysteresis cannot be explained using this law.

In a previous paper~\cite{Iwamatsu} we argued that the Cassie's law can be used 
to predict the {\it observed} contact angle only if the heterogeneity is of molecular 
size.  For macro- and micro-sized heterogeneity, we have derived, using the so-called 
interface displacement model~\cite{Swain}, the modified Cassie's law, which is similar 
but clearly different from the original Cassie's law. Our modified Cassie law predicts that the left and the right contact angle of the cylindrical droplet must be equal to minimize the total free energy.  Using this principle derived from the modified Cassie's law and the constraint on the volume, we can examine the behavior of the contact angle as the volume of the droplet is increased or decreased.

In this paper, we will discuss a simplified model for a translationally symmetric cylindrical droplet on a heterogeneously striped but smooth surface~\cite{Johnson}, and examine the hysteresis of the contact angle of the droplet as its volume is controlled externally.  Although a more realistic three-dimensional calculation has already been reported~\cite{Brandon,Lipowsky}, we believe that even a simplified model can provide a more profound and a new insight into the well-known hysteresis problem.

The format of this paper is as follows: In section2, we will present the minimum of modified Cassie's law~\cite{Iwamatsu} using a simplified model~\cite{Swain}. In section 3, we will use this modified Cassie's law as a vehicle to study the contact angle hysteresis of the liquid droplet on a chemically heterogeneous striped surface. 
Section 4 is devoted to the concluding remarks.

\section{The modified Cassie's law}

Since the modified Cassie's law was derived previously~\cite{Iwamatsu}, we will give a minimal form of the formula which is necessary to discuss the contact angle hysteresis.

\begin{figure}[htbp]
\begin{center}
\includegraphics[width=0.7\linewidth]{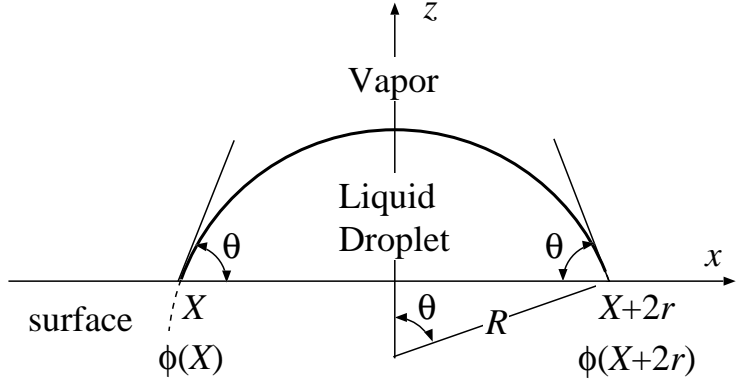}
\caption{A freely moving droplet on a heterogeneous surface.
The droplet can move freely along the $x$ coordinate.  
The left edge of the droplet is at $X$ and the right edge at $X+2r$.  
If the left and the right contact angle is different, the drop is 
dragged by the surface from lyophobic (less wettable with a higher 
contact angle) toward lyophilic (more wettable with a lower contact angle) part.
}
\label{Fig:1}
\end{center}
\end{figure}

We consider a cylindrical droplet of the base length $2r$ and the 
apparent contact angle $\theta$ with circular meniscus of radius 
$R=r/\sin\theta$ (Fig. \ref{Fig:1}) which can move 
freely along the heterogeneously striped surface.  The size of
the droplet considered is macroscopic so that the 
molecular-scale liquid-vapor interfacial width is smaller than the
scale of surface heterogeneity~\cite{Iwamatsu}, yet it is smaller than the capillary
length which is of the order of few mm so that the gravity is 
negligible~\cite{deGennes}.  

One should minimize the free energy of droplet given by
\begin{equation}
\mathcal{F} = \sigma_{lf}\frac{2r\theta}{\sin\theta}+\int_{X}^{X+2r}\left(\sigma_{sl}(x)-\sigma_{sf}(x)\right)dx \label{eq:3-1}
\end{equation}
where $\sigma_{lf}$, $\sigma_{sl}$, and $\sigma_{sf}$ are the
liquid-fluid (vapor), the solid (surface)-liquid, and solid-fluid surface
tensions respectively.  The left edge of the droplet is at $X$ and 
the right edge at $X+2r$.

Now, writing the surface free energy using the local contact 
angle $\phi(x)$ at $x$ defined by the Young's equation~\cite{Young}:
\begin{equation}
\sigma_{sl}(x)-\sigma_{sf}(x)=-\sigma_{lf}\cos\phi(x)\label{eq:3-2}
\end{equation}
we find
\begin{equation}
\mathcal{F} = \sigma_{lf}\frac{2r\theta}{\sin\theta}-\sigma_{lf}\int_{X}^{X+2r}\cos\phi(x)dx
\label{eq:3-3}
\end{equation}
which should be minimized with respect to the apparent contact angle $\theta$ 
and the position $X$ of the droplet (see Fig. \ref{Fig:1}) subject to the condition 
of the constant volume $S$ (which is actually the cross section):
\begin{equation}
S = \int_{-r}^{r}z(x)dx=r^{2}\frac{\theta-\sin\theta\cos\theta}{\sin^{2}\theta}
\label{eq:3-3x}
\end{equation}
which gives the relation between the base length $r$ and the contact angle $\theta$:
\begin{equation}
\frac{dr}{d\theta}=r\frac{\theta\cot\theta-1}{\theta-\sin\theta\cos\theta}
\label{eq:3-4}
\end{equation}

%\end{document}

By minimizing (\ref{eq:3-3}) with respect to $\theta$ and $r$ through (\ref{eq:3-4}), 
one obtains
\begin{equation}
\cos\theta = \cos\phi(X+2r)
\label{eq:3-6}
\end{equation}This equation should be augmented by the condition that 
the position $X$ of the droplet becomes the minimum of the free energy.  
This condition is derived by minimizing the above 
free energy (\ref{eq:3-3}) with respect to the position $X$, which leads to
\begin{equation}
\cos\phi(X+2r)=\cos\phi(X)
\label{eq:3-7}
\end{equation}

When the droplet can move freely, it will move to a position where the local 
contact angle $\phi(X)$ of the left edge and $\phi(X+2r)$ of the right edge 
become equal.  This condition has a simple mechanical meaning 
since eq. (\ref{eq:3-7}) can be written as
\begin{equation}
\sigma_{sf}(X+2r)-\sigma_{sl}(X+2r)= \sigma_{sf}(X)-\sigma_{sl}(X)
\label{eq:3-8}
\end{equation}
from Young's equation.  Since eq. (\ref{eq:3-8}) represents the balance of the net force exerted from the substrate at the left and the right edge of the droplet, the minimum total free energy implies the force balance.  This is a natural condition of static (non-moving) droplet.  Therefore a thermodynamic equilibrium is realized when the mechanical balance is satisfied~\cite{Iwamatsu}.  

When this condition is violated, the droplet can move freely.  For example, if $\phi(X+2r)<\phi(X)$, then 
\begin{equation}
\sigma_{sf}(X+2r)-\sigma_{sl}(X+2r)>\sigma_{sf}(X)-\sigma_{sl}(X)
\label{eq:3-9}
\end{equation}
Therefore the force exerted from the surface at the right edge $X+2r$ is stronger than the one at the left edge $X$, and the droplet moves to the right toward the area of lower contact angle.  Therefore, the droplet moves from the lyophobic part to the lyotropic part of the surface.  Such a force unbalance is widely used to create freely moving microscopic droplets~\cite{Chaundhury,Thiele}

Combining eqs.(\ref{eq:3-6}) and (\ref{eq:3-7}), we arrive at the modified Cassie's law:
\begin{equation}
\cos\theta = \frac{1}{2}\left(\cos\phi(X)+\cos\phi(X+2r)\right)
\label{eq:3-10}
\end{equation}
and eq. (\ref{eq:3-7}), which is similar, but completely different 
from the original Cassie's law~\cite{Cassie2}. Therefore, the apparent 
contact angle $\theta$ should be given by the 
local contact angle $\phi(X)$ at the left and $\phi(X+2r)$ at the right 
edge of the droplet.  All three must be equal for an equilibrium droplet. 
No averaging such as the original Cassie's law~\cite{Cassie2} can be 
used to predict the {\it observed} apparent contact angle unless 
the heterogeneity is of a nanoscopic molecular size~\cite{Iwamatsu,Zhang}.

\section{Contact angle hysteresis on periodically striped surface}

\begin{figure}[htbp]
\begin{center}
%\vspace*{13pt}
\includegraphics[width=0.8\linewidth]{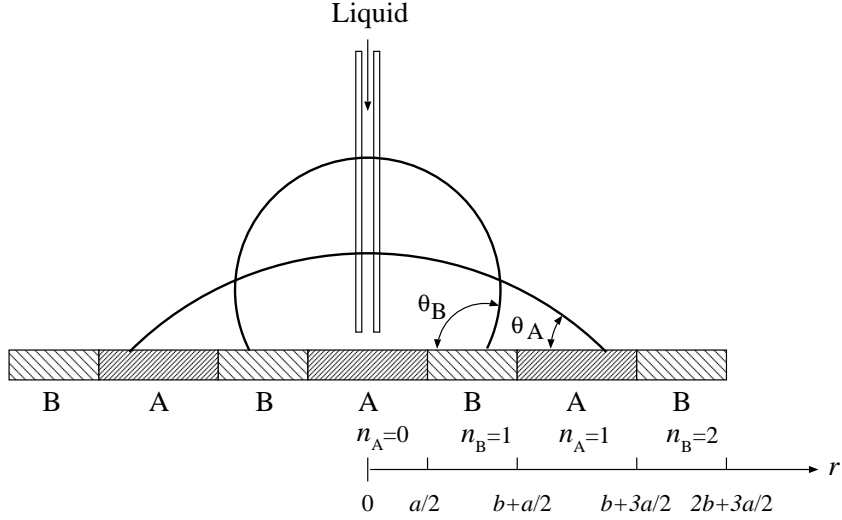}
%\vspace*{3cm}
\caption{A model heterogeneous striped surface which consists of alternating A 
and B surface with widths $a$ and $b$.  The droplet sits at the center.  
We assume without loss of generality that the A surface is at the 
center.  The droplet is characterized by the contact angle.  We call 
the droplet type-A when its contact angle is $\theta_{\rm A}$ and type-B 
when it is $\theta_{\rm B}$.  As we increase the size of the droplet, 
we may expect complex hysteresis to appear in order to satisfy the 
volume constraint. }
\label{Fig:2}
\end{center}
\end{figure}

From the modified Cassie's law (\ref{eq:3-7}) of the previous section, it is natural to assume that the droplet occupies a position where the local contact angle of the left edge $\phi(X)$ is equal to that of the right edge $\phi(X+2r)$.  Using this simple rule and a model of a cylindrical droplet on a periodically striped surface 
(Fig. \ref{Fig:2}), we will consider the contact angle hysteresis on a chemically heterogeneous striped surface.  A similar model with a sinusoidally modulated heterogeneous surface rather than a striped surface was examined~\cite{Marmur,Brandon,Brandon2}.  However, there is one important factor missing for such a continuous heterogeneity.  Namely the surface pinning at the interface of stripes, which could play a crucial role in various morphological phase transitions on the surface~\cite{Lipowsky}.  Furthermore, the striped surface would be more realistic than the sinusoidally modulated surface
considered by Marmur and coworkers~\cite{Marmur,Brandon,Brandon2}. 
Johnson and Dettre~\cite{Johnson} also considered a similar striped model 
but they considered an axisymmetric semi-spherical droplet.  Striped heterogeneity
was also considered by Neumann and Good~\cite{Neumann} but they only considered
the capillary rise along striped wall.

The heterogeneous surface model which we used in this study is depicted in Fig.~\ref{Fig:2}. The surface consists of two types of stripes that are periodically arranged.  The
stripes are aligned parallel to the axis of the cylindrical droplet. The surface "A" of width $a$ is characterized by the intrinsic contact angle $\theta_{\rm A}$, and "B" of width $b$ is characterized by $\theta_{\rm B}$.  At the AB interface, however, the droplet is pinned, and the apparent contact angel $\theta$ can take any values between $\theta_{\rm A}$ and $\theta_{\rm B}$~\cite{Lipowsky} to satisfy the volume constraint:
\begin{equation}
\theta_{\rm A}\leq \theta\leq \theta_{\rm B},\;\;\;\mbox{or}\;\;
\theta_{\rm A}\geq \theta\geq \theta_{\rm B}
\label{eq:4-1}
\end{equation}
The origin of the coordinate is chosen at the center of one of the stripes of type "A".

In order to make the apparent contact angle of the left and the 
right edge equal, it is natural to assume that the center of droplet 
is fixed at the origin and always occupies the symmetric position 
(Fig.\ref{Fig:2}).  We will call the droplet type-A if its left and right 
contact angles are $\theta_{\rm A}$ , and type-B if they are 
$\theta_{\rm B}$.

\begin{figure}[htbp]
\begin{center}
%\vspace*{13pt}
\includegraphics[width=0.7\linewidth]{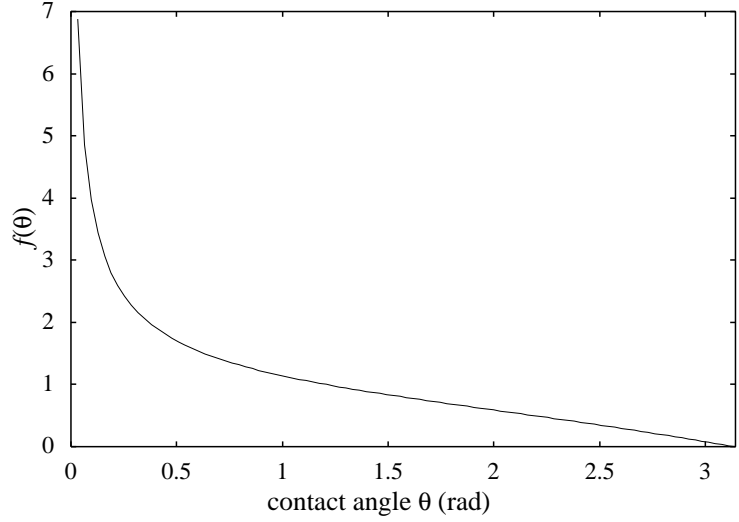}
%\vspace*{3cm}
\caption{The coefficient $f(\theta)$ in (\ref{eq:4-2}) as the 
function of the contact angle $\theta$. It approaches $0$ as 
$\theta\rightarrow \pi$, and it diverges as $\theta \sim 1/\sqrt{\theta}$ as 
$\theta\rightarrow 0$. }
\label{Fig:3}
\end{center}
\end{figure}

The position $r$ of the edge of droplet as the function of droplet 
volume $S$ is given by
\begin{equation}
r=\sqrt{\frac{\sin^{2}\theta_{i}}{\theta_{i}-\sin\theta_{i}\cos\theta_{i}}}S^{1/2}=f(\theta_{i})S^{1/2},\;\;\;\;\;(i={\rm A}, {\rm B})
\label{eq:4-2}
\end{equation}
from (\ref{eq:3-3x}), where $\theta_{i}$ is the intrinsic contact angle of surface $i=$A and $i=$B.  
Figure \ref{Fig:3} shows the coefficient $f(\theta)$ as the function of 
the contact angle $\theta$. 

As we increase the volume $S$ of the droplet, the edge $r$ of the droplet moves according to (\ref{eq:4-2}).  As the edge reaches the AB interface, the edge is pinned, and the apparent contact angel $\theta$ is determined from the droplet volume as far as the angle $\theta$ can satisfies eq. (\ref{eq:4-1}).  From now on, we will call (AB) interface when the droplet edge moves from A to B surface and (BA) interface when the droplet edge moves from B to A surface on increasing the size of droplet.  At these two types of interface, we may expect a sudden jump of the droplet edge by a finite length or the continuous changes of contact angle on increasing or decreasing the droplet volume.

The free energy of the droplet when its edge is at $r$ is given by
\begin{equation}
\frac{F_{\rm A}}{\sigma_{lf}}= 2r\left(\frac{\theta}{\sin\theta}-\cos\theta_{\rm A}\right)
-2n_{\rm A}b\left(\cos\theta_{\rm B}-\cos\theta_{\rm A}\right),\;\;(n_{\rm A}=0,1,\dots)
\label{eq:4-3}
\end{equation}
when the edge of the droplet is on $n_{\rm A}$-th "A"-type surface, 
where the index $n_{\rm A}$ is counted from the origin (Fig. \ref{Fig:2}).  
The apparent contact angle is $\theta=\theta_{\rm A}$ if the edge is 
not pinned at the (AB) or (BA) interfaces.  If it is pinned, $\theta$ 
can take any value between $\theta_{\rm A}$ and $\theta_{\rm B}$ as 
in (\ref{eq:4-1}).

When the edge of the droplet is on $n_{\rm b}$-th "B"-type surface, 
the free energy is given by
\begin{eqnarray}
\frac{F_{\rm B}}{\sigma_{lf}}= 2r\left(\frac{\theta}{\sin\theta}-\cos\theta_{B}\right)
-(2n_{\rm B}-1)a\left(\cos\theta_{A}-\cos\theta_{B}\right),\;
(n_{\rm B}=1,2,\dots) \nonumber \\
\label{eq:4-4}
\end{eqnarray}
The free energy that is necessary to increase the droplet volume 
can be calculated from these two expressions (\ref{eq:4-3}) and (\ref{eq:4-4}).

\begin{figure}[htbp]
\begin{center}
%\vspace*{13pt}
\includegraphics[width=0.7\linewidth]{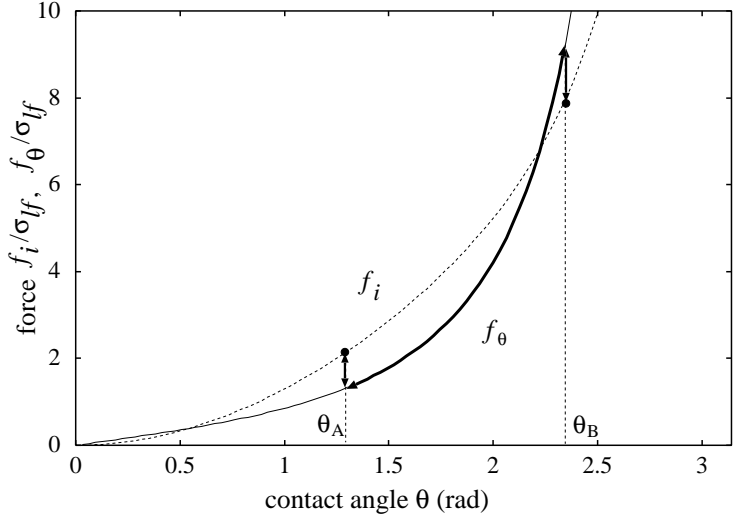}
%\vspace*{3cm}
\caption{
The force $f_{i}$ (\ref{eq:4-5}) and $f_{\theta}$ (\ref{eq:4-6}) are functions of the contact angle $\theta$. They are always positive with the increasing function of $\theta$. The force increases along the curve $f_{\theta}$ from $\theta_{\rm A}$ to $\theta_{\rm B}$ if $\theta_{\rm A}<\theta_{\rm B}$ when the droplet edge is pinned at the AB interface.  The jump in the force is expected when the edge is unpinned when $\theta=\theta_{\rm A}$ and $\theta=\theta_{\rm B}$.
}
\label{Fig:4}
\end{center}
\end{figure}

The force $f_{i}$ necessary to increase the volume of the droplet 
on the "$i$"-type surface is given by
\begin{equation}
f_{i}=\frac{\partial F_{i}}{\partial r}=2\left(\frac{\theta}{\sin\theta}-\cos\theta\right)\sigma_{lf},\;\;\;\;\;(\theta=\theta_{\rm A}, \theta_{\rm B})
\label{eq:4-5}
\end{equation}
Similarly, the force $f_{\theta}$ necessary to increase the volume when 
the droplet edge is pinned at the strip interface is given by
\begin{equation}
f_{\theta}=\frac{\partial F_{i}}{r\partial \theta}
=2\left(\csc\theta-\theta\cot\theta\csc\theta\right)\sigma_{lf}
\label{eq:4-6}
\end{equation}
They are all positive and increasing functions of the contact angle $\theta$ as shown in Fig. \ref{Fig:4}.  Then a stronger (constant) force is necessary to increase the volume of the droplet when droplet is on the surface that is not wet and has a larger contact angle.  When the contact lines are pinned at the AB interface, the force increases steadily as we increase the contact angle $\theta$ along the curve $f_{\theta}$.
The jump in the force (see Fig. \ref{Fig:4}) is expected when the droplet edge is unpinned and leaves the interface when $\theta=\theta_{\rm A}$ and $\theta=\theta_{\rm B}$.

%\end{document}

\begin{figure}[htbp]
\begin{center}
\includegraphics[width=1.0\linewidth]{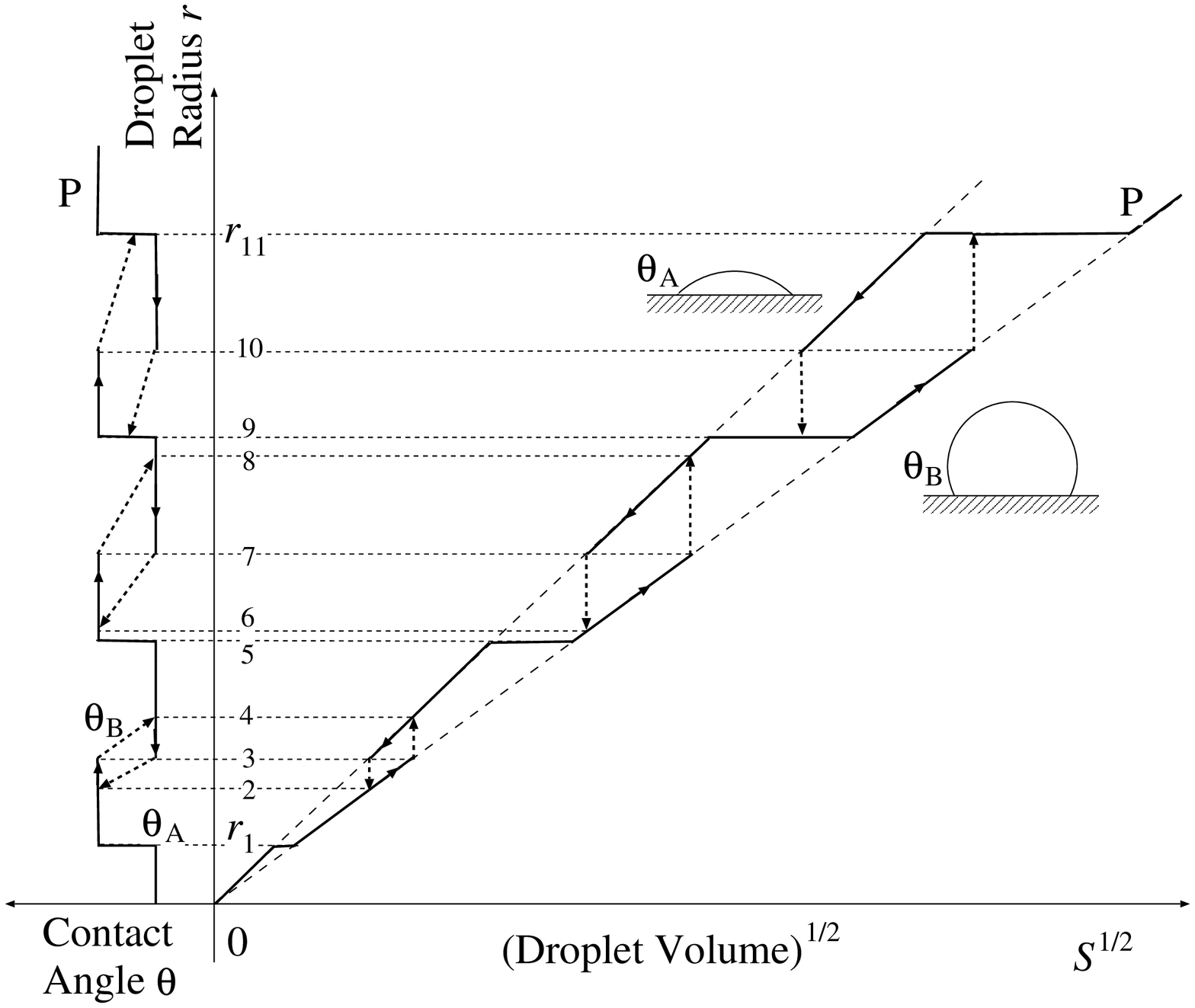}
\caption{A hysteresis of the contact angle and the 
droplet morphology as we increase and decrease 
the volume $S$ of the droplet 
when $\theta_{\rm A}<\theta_{\rm B}$.  }
\label{Fig:5}
\end{center}
\end{figure}

In figure \ref{Fig:5} we show two straight lines (\ref{eq:4-2}) when $\theta_{A}\leq \theta_{B}$.  As we increase the volume $S$ of the smallest type-A droplet, the contact angle remains $\theta_{A}$ (type-A droplet) until the droplet edge touches the (AB) interface at $r_{1}=a/2$ (Fig.\ref{Fig:5}). Then the contact angle $\theta$ starts to increase continuously from $\theta_{A}$ to $\theta_{B}$ while the edge is pinned at the (AB) interface at $r_{1}$.  When the contact angle becomes $\theta_{B}$, the droplet becomes type-B and its edge starts to move toward outside until it reaches the (BA) interface at $r_{3}=b+a/2$.  As we increase the volume $S$ further, its edge jumps from the (BA) interface at $r_{3}$ to $r_{4}$ $(b+a/2\leq r_{4} \leq b+3a/2)$ and the liquid droplet transforms into type-A and the contact angle decreases to $\theta_{A}$. The free energy change $\Delta F_{4}$ from type-B to type-A droplet can be calculated as
\begin{equation}
\frac{\Delta F_{4}}{\sigma_{lf}}=2r_{4}\frac{\theta_{A}}{\sin\theta_{A}}
-(a+2b)\frac{\theta_{B}}{\sin\theta_{B}}-(2r_{4}-a-2b)\cos\theta_{A}
\label{eq:4-7}
\end{equation}
If it is positive, the free energy increases and the external work is necessary 
to release the droplet edge from the (BA) interface at $r_{3}$. If it 
is negative, the interface moves spontaneously.

%\end{document}
As we increase the volume of this type-A droplet further the 
droplet edge reaches the (AB) interface at $r_{5}=b+3a/2$, then 
the contact angle $\theta$ increases continuously until 
the droplet becomes type-B with the contact angle $\theta_{\rm B}$, 
while the edge is pinned at $r_{5}$.  The work $\Delta F_{5}$ that is 
necessary during this process is
\begin{equation}
\frac{\Delta F_{5}}{\sigma_{lf}}
=(3a+2b)\left(\frac{\theta_{B}}{\sin\theta_{B}}-\frac{\theta_{A}}{\sin\theta_{A}}\right)
\label{eq:4-8}
\end{equation}
which is always positive as $\theta_{\rm B}\geq \theta_{\rm A}$.

In this way, the type-A and type-B droplets appear one after 
another until the droplet grows sufficiently large.  After the 
droplet edge reaches the (BA) interface at $r_{9}=2b+5a/2$, only the 
type-B droplet appears on increasing the droplet volume.

If we increase the droplet volume further, the type-A droplet
cannot appear any longer because the droplet volume is so large that
the droplet of the same volume with smaller contact angle $\theta_{A}$ 
cannot exist. When the droplet edge of the type-B droplet 
reaches the (BA) interface as we increase the volume, the droplet
edge jumps to the outer (AB) interface and makes a contact angle 
$\theta$ ($\theta_{A}<\theta<\theta_{B}$).  If we further 
increase the volume, the contact angle
starts to increase until it becomes $\theta_{B}$ while the droplet
edge is pinned at the (AB) interface.  In this way, the type-A droplet
cannot appear above certain droplet volume and only the type-B droplet appears.  
The contact angle of this type-B droplet, however, can be smaller than
$\theta_{B}$ when its edge is pinned at the (AB) interface.  Of course
this argument is valid only when the size of the droplet is smaller than
the capillary length since our formulation assumes semi-circular
meniscus and neglects the effect of gravity.

Next, we consider the reverse process.  We start from a 
sufficiently large type-B droplet at P in Fig.~\ref{Fig:5}.  
As we decrease the volume $S$, the contact angle starts to decrease 
from $\theta_{B}$ toward $\theta_{A}$ at (AB) interface at $r_{11}=3b+7a/2$, 
while its edge is pinned.  As the volume decreases further, the 
droplet transforms into type-A and starts to shrink.  
As its edge reach the (BA) interface at $r_{10}=3b+5a/2$, the 
edge jumps to the next (AB) interface 
at $r_{9}=2b+5a/2$ and continues to shrink as a type-A droplet.  
In this reverse process, the type-B 
droplet appears only in the interval $(r_{5},r_{6})$ 
and $(r_{1},r_{2})$, where $b+3a/2\leq r_{6} \leq 2b+3a/2$ 
and $a/2\leq r_{2} \leq b+a/2$. 

Naturally the type-B droplet cannot appear if the droplet
volume is sufficiently large in this case.  
When the droplet edge of the type-A droplet with sufficiently
large volume reaches the (BA) interface as we decrease the volume, 
the droplet edge jumps to the inner (AB) interface and makes a 
contact angle $\theta$ ($\theta_{A}<\theta<\theta_{B}$).  If we 
further decrease the volume, the contact angle
starts to decrease until it becomes $\theta_{A}$ while the droplet
edge is pinned at the (AB) interface.  In this way, the type-B droplet
cannot appear above certain droplet volume and only the type-A 
droplet appears in this reverse process.  
The contact angle of this type-A droplet, however, can be larger than
$\theta_{A}$ when its edge is pinned at the (AB) interface.

\begin{figure}[htbp]
\begin{center}
%\vspace*{13pt}
\includegraphics[width=0.6\linewidth]{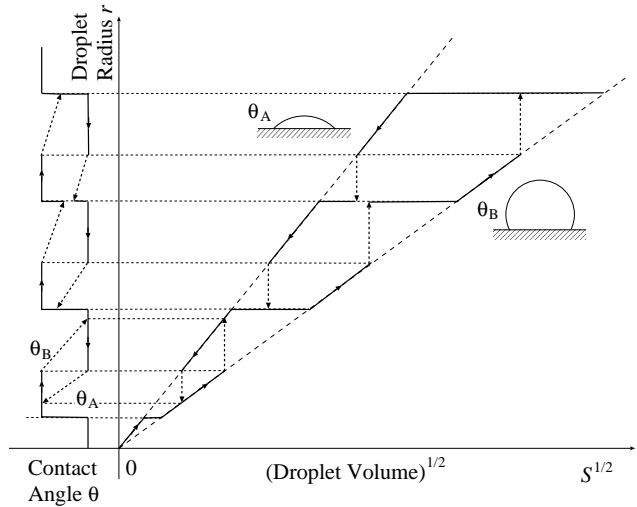}
%\vspace*{3cm}
\caption{The same as Fig. \ref{Fig:5} when the contact angle 
$\theta_{A}$ is much smaller than $\theta_{B}$.
}
\label{Fig:6}
\end{center}
\end{figure}

From the above argument, a complicated hysteresis 
will be expected even for this simple striped surface.  
On 
increasing the volume, the type-B droplet with larger contact 
angle $\theta_{B}$ ($>\theta_{A}$) appears 
predominantly, while the type-A droplet with a smaller contact angle $\theta_{A}$ 
appears predominantly when 
we decrease the volume.  A
stronger hysteresis is expected if $\theta_{B}$ is much larger than $\theta_{A}$, then the type-B droplet 
with larger contact angle appears more predominantly on 
increasing the volume while the type-A droplet appears 
on decreasing the volume (Fig.\ref{Fig:6}).  Therefore, the 
advancing contact angle is predominantly the larger one 
while the receding one is predominantly the smaller one 
unless the droplet is too small, which is usually observed 
experimentally~\cite{Butt,Drelich}.

\begin{figure}[htbp]
\begin{center}
%\vspace*{13pt}
\includegraphics[width=0.6\linewidth]{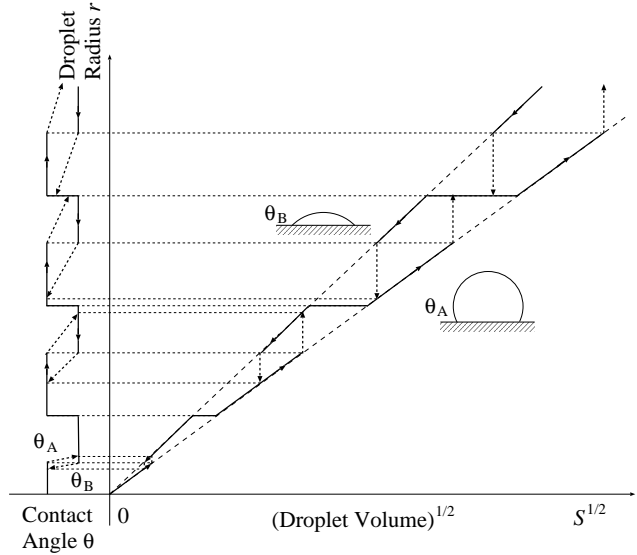}
%\vspace*{3cm}
\caption{The same as Fig. \ref{Fig:5} when the contact angle 
$\theta_{A}$ is much larger than $\theta_{B}$.
}
\label{Fig:7}
\end{center}
\end{figure}

If the contact angle $\theta_{A}$ is larger than $\theta_{B}$, the story is 
slightly different only for the smallest droplet.  As shown in Fig. \ref{Fig:7}, 
the initial loop of the hysteresis is different, but the story is essentially 
the same as in Fig. \ref{Fig:6} and Fig. \ref{Fig:7} for larger droplet.  
In this case, the droplet 
with larger contact angle is type-A and the smaller one is type-B.   Therefore, the 
conclusion drawn from Fig. \ref{Fig:5} and \ref{Fig:6} applies:  On increasing the 
droplet volume, the droplet with a larger contact angle appears, while smaller 
one appears on decreasing the volume.

\section{Concluding Remarks}
\label{sec:sec4}

In this paper, we used a simplified cylindrical model 
of a droplet on a chemically heterogeneous smooth surface to derive the 
expression for the apparent contact angle similar to but 
different from the Cassie's law.  From this modified 
Cassie law, the local contact angle of the left and the right edges 
of the droplet should be equal for the thermodynamically equilibrium 
droplet.  This condition express the force balance acting 
at the edge of the droplet from the substrate.  Thus, for 
a static droplet, the local contact angle of the left and 
the right edge must be equal. This force balance condition 
explains moving and driving a micro-droplet along the surface.

Using the above condition that the contact angles at the left and the right edge must be equal, we have examined the behavior of the droplet on a  periodically striped surface composed of A and B surfaces when we control the volume of droplet.  It has been shown that the complex hysteresis is expected on such a striped surface.  For example the edge of the droplet jump to a finite width or the contact edge is pinned at (AB) and (BA) interface and the contact angle changes continuously.  It is generally expected that only the droplet with a larger contact angle appears when we increase the volume of droplet, while only the droplet with smaller contact angle appears when we decrease the volume. 
Therefore, our model is consistent with the observation that the
advancing contact angle is usually higher
than the receding contact angle~\cite{Butt,deGennes}.
Given the recent advancement of the nanomaterial, such a striped surface could be achieved experimentally and the hysteresis of the contact angle could also be observed.

In our report, we used the argument based on the free energy
of droplet with cylindrical shape.  The free energy barrier and the vibration
state due to the deformation of the droplet discussed by Johnson and Dettre~\cite{Johnson} are
beyond the scope of the present work, which could play some role during the
jumping motion of the droplet edge.
We also restricted our discussion to the macro- and microscopic droplet, and the macroscopic apparent contact angle is considered, which could be observed using a microscope.  In order to study the nanoscopic contact angle of a nanoscopic droplet~\cite{Dietrich2}, which could be observed, for example, using an atomic force microscope (AFM)~\cite{Butt}, we have to include the thin film potential.  These issues are left for future theoretical as well as experimental investigations.

\end{document}